\documentclass[journal,12pt,onecolumn]{IEEEtran}

\usepackage{setspace}
\ifCLASSOPTIONonecolumn \doublespace \fi
\usepackage{amsfonts}
\usepackage{amsmath}
\usepackage{amsthm}
\usepackage{graphicx}
\usepackage{cite}
\allowdisplaybreaks[1]

\graphicspath{{Figures/}} \DeclareGraphicsExtensions{.eps,.ps}

\title{On the Matrix Inversion Approximation Based on Neumann Series in Massive MIMO Systems}
\author{\IEEEauthorblockN{Dengkui Zhu\IEEEauthorrefmark{1}, Boyu Li\IEEEauthorrefmark{1}, and Ping Liang\IEEEauthorrefmark{1}\IEEEauthorrefmark{2}}
\\
\IEEEauthorblockA{\IEEEauthorrefmark{1}RF DSP Inc., 30 Corporate Park, Suite 210, Irvine, CA 92606, USA,\\ e-mail: dkzhu@rfdsp.com, byli@rfdsp.com, pliang@rfdsp.com}
\IEEEauthorblockA{\IEEEauthorrefmark{2}Department of Electrical Engineering, University of California - Riverside, Riverside, CA 92521, USA,\\ e-mail: liang@ee.ucr.edu} 
}
\begin{document}


\maketitle
\begin{abstract}
Zero-Forcing (ZF) has been considered as one of the potential practical precoding and detection method for massive MIMO systems. One of the most important advantages of massive MIMO is the capability of supporting a large number of users in the same time-frequency resource, which requires much larger dimensions of matrix inversion for ZF than conventional multi-user MIMO systems. In this case, Neumann Series (NS) has been considered for the Matrix Inversion Approximation (MIA), because of its suitability for massive MIMO systems and its advantages in hardware implementation. The performance-complexity trade-off and the hardware implementation of NS-based MIA in massive MIMO systems have been discussed. In this paper, we analyze the effects of the ratio of the number of massive MIMO antennas to the number of users on the performance of NS-based MIA. In addition, we derive the approximation error estimation formulas for different practical numbers of terms of NS-based MIA. These results could offer useful guidelines for practical massive MIMO systems.      
\end{abstract}

\section{Introduction} \label{sec:introduction}
Massive Multiple-Input Multiple-Output (MIMO) systems were firstly introduced in \cite{Marzetta_massive_MIMO_original}, and have drawn great interest form both academia and industry. In such systems, each Base Station (BS) is equipped with dozens to hundreds of antennas to serve tens of users in the same time-frequency resource. Therefore, they can achieve significantly higher spatial multiplexing gains than conventional multi-user MIMO systems, which offers one of the most important advantages of massive MIMO systems, the potential capability to offer linear capacity growth without increasing power or bandwidth \cite{Marzetta_massive_MIMO_original,Rusek_massive_MIMO_overview,Larsson_massive_MIMO_overview, Hoydis_massive_MIMO}.
 
It has been shown that, for massive MIMO systems where the number of antennas $M$, e.g., $M=128$, is much larger than the number of served users $K$, e.g., $K=16$, \cite{Hoydis_massive_MIMO,Rusek_massive_MIMO_overview}, Zero-Forcing (ZF) precoding and detection can achieve performance very close to the channel capacity for the downlink and uplink respectively \cite{Rusek_massive_MIMO_overview}. As a result, ZF has been considered as one of the potential practical precoding and detection method for massive MIMO systems \cite{Shepard_Argos,Hoydis_massive_MIMO,Rusek_massive_MIMO_overview,Marzetta_ZF_CB}. 

For the hardware implementation of ZF, despite of the very large number of $M$, the main complexity is the inverse of a $K\times K$ matrix \cite{Rusek_massive_MIMO_overview,Prabhu_Neumann,Wu_Neumann}. Unfortunately, for massive MIMO systems, although $K$ is much smaller than $M$, it is still much larger than conventional multi-user MIMO systems. As a result, in this case, the computation of the exact inversion of the $K\times K$ matrix could result in very high complexity \cite{Wu_Neumann}, which may cause large processing delay so that the demands of the channel coherence time is not met. Due to this reason, Neumann Series (NS) has been considered to carry out the Matrix Inversion Approximation (MIA), because it is well suited for massive MIMO systems and it is advantageous for hardware implementation \cite{Rusek_massive_MIMO_overview,Prabhu_Neumann,Wu_Neumann}. 

Despite of the advantages, some potential application issues of the NS-based MIA have also been identified. Firstly, for a finite $M/K$ ratio, the NS may not converge, resulting the failure of the algorithm \cite{Rusek_massive_MIMO_overview,Prabhu_Neumann}. What $M/K$ ratio could offer high convergence probability is still not clear. Secondly, for the NS-based MIA to achieve good performance with quick convergence, the $K\times K$ matrix needs to be diagonally dominant \cite{Wu_Neumann}. In order to satisfy this condition, $M\gg K$ is required \cite{Rusek_massive_MIMO_overview,Wu_Neumann}. Similarly, what $M/K$ ratio could provide high probability of diagonally dominant is also not clear. Moreover, with a larger number of terms, the NS-based MIA offers closer performance to the exact inversion \cite{Wu_Neumann,Prabhu_Neumann}. However, the larger number of terms results in more processing cycles. Hence, for practical hardware implementation, the number of terms cannot be very large. Although the approximation error analysis was carried out and a residual error upper bound of the NS-based MIA was derived\cite{Wu_Neumann}, the approximation error analysis with high accuracy has not been derived.

In this paper, we address the three problems listed above. Specifically, we firstly derived a $M/K$ ratio condition that offers high convergence probability. Then, we derived another $M/K$ ratio condition that provides high probability for the $K\times K$ matrix to be diagonally dominant. Finally, we carry out the approximation error analysis with high accuracy for practical numbers of terms for the NS-based MIA in hardware implementation.

The remainder of this paper is organized as follows. In Section \ref{sec:basis}, the basis of the NS-based MIA in massive MIMO systems is briefly reviewed. The $M/K$ ratio condition that provides high convergence probability is derived in Section \ref{sec:convergence}. Then, another $M/K$ ratio condition that offers high diagonally dominant probability for the $K\times K$ matrix is derived in Section \ref{sec:dominant}. In Section \ref{sec:error}, the approximation error analysis with high accuracy for practical numbers of terms for the NS-based MIA is carried out. Finally, after a discussion in Section \ref{sec:discussion}, conclusions are drawn in Section \ref{sec:conclusions}.

 
\section{Basis of NS-Based MIA in Massive MIMO Systems} \label{sec:basis}
Consider a massive MIMO wireless system where the BS is equipped with $M$ antennas to serve $K$ single-antenna users in the same time-frequency resource. Then, for the uplink, the $M\times K$ channel matrix is represented by $\mathbf{H} = \left[ h_{mk} \right]$, where $h_{mk}$ denotes the channel coefficient between the $m$th antenna and the $k$th user, with $m=1,\ldots,M$, and $k=1,\ldots,K$. Similarly to \cite{Rusek_massive_MIMO_overview,Prabhu_Neumann,Wu_Neumann}, the analysis in this paper assumes that the $h_{mk}$ elements are in uncorrelated Rayleigh flat fading, i.e., independent and identically distributed (i.i.d.) zero-mean unit-variance complex Gaussian variables. Note that, for the Time-Division Duplexing (TDD) mode, due to the channel reciprocity, the downlink has the same channel matrix $\mathbf{H}$ as the uplink, as long as the transmission duration is within the channel coherence time \cite{Marzetta_massive_MIMO_original,Rusek_massive_MIMO_overview,Larsson_massive_MIMO_overview, Hoydis_massive_MIMO,Shepard_Argos,Marzetta_ZF_CB}.  

In order to carry out ZF precoding for the downlink or the ZF detection for the uplink, the pseudo-inverse of $\mathbf{H}$ needs to be calculated \cite{Shepard_Argos,Hoydis_massive_MIMO,Rusek_massive_MIMO_overview,Marzetta_ZF_CB}, which is written as 
\begin{equation}
\mathbf{H}^\dag = \left( {\mathbf{H}^\mathrm{H} \mathbf{H}} \right)^{-1} \mathbf{H}^\mathrm{H}.
\label{eq:inv}
\end{equation}
Let $\mathbf{G} = \mathbf{H}^\mathrm{H} \mathbf{H} $. Then, in (\ref{eq:inv}), despite of the very large number of $M$, e.g., $256$, in massive MIMO systems, the main complexity of the hardware implementation lies in the inversion of the $K\times K$ matrix $\mathbf{G}$ \cite{Rusek_massive_MIMO_overview,Prabhu_Neumann,Wu_Neumann}. To exploit the large spatial multiplexing gains of massive MIMO systems, although much smaller than $M$, the number of $K$ is much larger than conventional multi-user MIMO systems, e.g., $K=16$. As a result, the complexity of calculating $\mathbf{G}^{-1}$ may be too high for hardware implementation. To address this issue, NS has been considered to carry out the MIA, because it is advantageous in hardware implementation and it is suitable for massive MIMO systems \cite{Rusek_massive_MIMO_overview,Prabhu_Neumann,Wu_Neumann}. Specifically, it can be written as   
\begin{equation}
\mathbf{G}^{-1}_{N} \approx \sum_{n=0}^{N-1} \left( \mathbf{I}_K - \mathbf{\Theta} \mathbf{G} \right)^n \mathbf{\Theta},
\label{eq:neumann_series}
\end{equation}
where $N$ denotes the number of terms used in the NS, and $\mathbf{\Theta}$ is a $K\times K$ diagonal matrix. Note that for (\ref{eq:neumann_series}) to work, the requirement below has to be satisfied 
\begin{equation}
\lim_{n\rightarrow \infty} \left( {\mathbf{I}_K - \mathbf{\Theta} \mathbf{G}} \right)^n \rightarrow \mathbf{0}_K.
\label{eq:neumann_condition}
\end{equation}

Note that $\mathbf{G}$ is a complex central Wishart matrix because the elements of $\mathbf{H}$ are i.i.d. complex Gaussian random variables \cite{Horn_Matrix_Analysis}. Let $\alpha = M/K$. As $K$ and $M$ grow, 
as derived in \cite{Alan_Eig_Wishart}, the largest and the smallest eigenvalues of $\mathbf{G}$ converge respectively to
\begin{align}
\lambda_{\mathrm{max}}\left( \mathbf{G} \right) \rightarrow M \left(1 + {1 \over \sqrt{\alpha}} \right)^2, \nonumber \\ 
\lambda_{\mathrm{min}}\left( \mathbf{G} \right) \rightarrow M \left(1 - {1 \over \sqrt{\alpha}} \right)^2. 
\label{eq:eigenvalues}
\end{align}
As a result, if $\mathbf{\Theta}$ is chosen as \cite{Rusek_massive_MIMO_overview}, which is
\begin{equation}
\mathbf{\Theta} = {\alpha \over M \left(1+\alpha \right)} \mathbf{I}_K = {1 \over M + K} \mathbf{I}_K,
\label{eq:diagonal_choice1}
\end{equation}
then, 
\begin{align}
\lambda_{\mathrm{max}}\left( \mathbf{\Theta} \mathbf{G} \right) \rightarrow 1 + {2 \sqrt{\alpha} \over 1 + \sqrt{\alpha}}, \nonumber \\
\lambda_{\mathrm{min}}\left( \mathbf{\Theta} \mathbf{G} \right) \rightarrow 1 - {2 \sqrt{\alpha} \over 1 + \sqrt{\alpha}}.
\label{eq:eigenvalues_choice1}
\end{align}
Therefore, the eigenvalues of $(\mathbf{I}_K - \mathbf{\Theta} \mathbf{G})$ lie approximately in the range of $[ -2\sqrt{\alpha}/(1+\alpha), 2\sqrt{\alpha}/(1+\alpha) ]$ \cite{Rusek_massive_MIMO_overview,Prabhu_Neumann}. Since $2\sqrt{\alpha}/(1+\alpha) \leq 1$ when $\alpha \geq 1$, the convergence of (\ref{eq:neumann_condition}) is satisfied with the choice (\ref{eq:diagonal_choice1}). Moreover, when $\alpha$ is very large, $2\sqrt{\alpha}/(1+\alpha) \rightarrow 0$, which means that (\ref{eq:neumann_condition}) converges very quickly. Hence, a small number of $N$ in (\ref{eq:neumann_series}) can offer close performance to the exact inverse.

Unfortunately, for finite $M$ and $K$ values, the eigenvalues of the product $\mathbf{\Theta} \mathbf{G}$ for a particular channel realization can lie outside the range of $[ -2\sqrt{\alpha}/(1+\alpha), 2\sqrt{\alpha}/(1+\alpha) ]$ \cite{Rusek_massive_MIMO_overview,Prabhu_Neumann}, which results in the failure of (\ref{eq:neumann_condition}). To address this issue, an attenuation factor $\delta$ where $0<\delta<1$ was introduced in \cite{Rusek_massive_MIMO_overview}, so (\ref{eq:diagonal_choice1}) changes to
\begin{equation}
\mathbf{\Theta} = {\delta \over M + K} \mathbf{I}_K.
\label{eq:diagonal_choice2}
\end{equation}
However, the proper choice of $\delta$ is hard to be determined. On the one hand, if $\delta$ is too large, the non-convergence issue still exists. On the other hand, if $\delta$ is too small, the convergence speed becomes very slow, so the number of $N$ needs to be very large to offer a good MIA, increasing the burden of the hardware implementation. 

Instead of (\ref{eq:diagonal_choice2}), another $\mathbf{\Theta}$ was applied in \cite{Prabhu_Neumann,Wu_Neumann}, which achieves a better MIA \cite{Prabhu_Neumann}. Specifically, $
\mathbf{G}$ is decomposed as 
\begin{equation}
\mathbf{G} = \mathbf{D} + \mathbf{E},
\label{eq:decompose}
\end{equation}
where $\mathbf{D}$ is a diagonal matrix including the diagonal elements of $\mathbf{G}$, and $\mathbf{E}$ is a hollow matrix including the off-diagonal elements of $\mathbf{G}$. Then, $\mathbf{\Theta}$ is chosen as
\begin{equation}
\mathbf{\Theta} = \mathbf{D}^{-1}.
\label{eq:diagonal_choice3}
\end{equation}
To achieve a good MIA with quick convergence, (\ref{eq:diagonal_choice3}) requires that $\mathbf{G}$ is a Diagonally Dominant Matrix (DDM) \cite{Prabhu_Neumann,Wu_Neumann}, i.e., 
\begin{equation}
|g_{ii}| > \sum_{j, j\neq i} |g_{ij}|, i,j=1,\ldots,K. 
\label{eq:dominant_condition1}
\end{equation}
The performance-complexity trade-off and hardware implementation of the NS-based MIA employing (\ref{eq:diagonal_choice3}) have been discussed for the downlink and uplink in \cite{Prabhu_Neumann} and \cite{Wu_Neumann} respectively. In both cases, the NS-based MIA employing (\ref{eq:diagonal_choice3}) was considered as a promising and practical method for massive MIMO systems. As a result, the analysis carried out in this paper is based on the choice of (\ref{eq:diagonal_choice3}). 

As mentioned in Section \ref{sec:introduction}, there still some issues on the application of (\ref{eq:diagonal_choice3}) for finite $M$ and $K$ values. Firstly, it is unclear that what $\alpha$ can offer high convergence probability. Secondly, it is unclear that what $\alpha$ can achieve high probability for $\mathbf{G}$ to be diagonal dominant. Moreover, more accurate approximation error analysis for practical $N$ values is needed. In the next sections, the aforementioned issues are addressed.
\section{Convergence and $\alpha$}\label{sec:convergence} 
According to the theory of matrix power series \cite{Horn_Matrix_Analysis}, for a $K \times K$ matrix $\mathbf{B}$, the product $\mathbf{B}^N$ converges to $\mathbf{0}_K$ only when the spectral radius of $\mathbf{B}$, denoted by $\rho(\mathbf{B})$, i.e., the maximum modulus of eigenvalues of $\mathbf{B}$, is less than $1$. Then, for the choice of (\ref{eq:diagonal_choice3}), a good MIA of (\ref{eq:neumann_series}) requires
\begin{equation}
\rho \left( \mathbf{I}_K - \mathbf{D}^{-1} \mathbf{G} \right) < 1.
\label{eq:convergence_condition1}
\end{equation}
Since the elements of $\mathbf{H}$ are i.i.d. zero-mean unit-variance complex Gaussian random variables, when the number of $M$ is large, the diagonal elements of $\mathbf{D}$ approach to $M{\mathrm{E}}\{ | {h_{kk} |^2 } \} = M$ by the law of large numbers \cite{Marzetta_massive_MIMO_original,Rusek_massive_MIMO_overview, Hoydis_massive_MIMO}. Therefore, the diagonal matrix $\mathbf{D}$ can be replaced by $M\mathbf{I}_K,$ Then, the condition (\ref{eq:convergence_condition1}) changes to
\begin{equation}
\left| M - \lambda \left( {\mathbf{G}} \right) \right| < M \Rightarrow 0 < \lambda \left( {\mathbf{G}} \right) < 2M. 
\label{eq:convergence_condition2}
\end{equation}
As $\mathbf{G}$ is a positive-definite matrix \cite{Horn_Matrix_Analysis}, its eigenvalues are all larger than $0$ \cite{Horn_Matrix_Analysis}. As a result, (\ref{eq:convergence_condition2}) is equivalent to 
\begin{equation}
\lambda_{\mathrm{max}} ( \mathbf{G} ) < 2M.
\label{eq:convergence_condition3}
\end{equation}

Note that $\mathbf{G}=\mathbf{H}^\mathrm{H}\mathbf{H}$ is a complex central Wishart matrix \cite{Horn_Matrix_Analysis}, and 
the distribution of $\lambda_{\mathrm{max}}(\mathbf{G})$ is provided in \cite{Ratnarajah_Eig_Wishart} as
\begin{align}
P\left( \lambda_{\mathrm{max}}(\mathbf{G}) < x \right) = & \frac{\mathcal{C}\Gamma_K \left( K \right)}{\mathcal{C}\Gamma_K \left( M + K \right)} x^{KM} \nonumber \\ 
& \times {}_1F_1 \left(M;M + K; - x{\mathbf{I}} \right),
\label{eq:eigen_largest}
\end{align}
where $x$ is a non-negative number. The complex multivariate gamma function $\mathcal{C}\Gamma_p(a)$ is defined as 
\begin{equation}
\mathcal{C}\Gamma_p\left(a\right) = \pi^{p\left(p-1\right)/2} \prod_{i=1}^p \Gamma \left[a - i +1 \right],
\label{eq:complex_multi_gamma}
\end{equation}
where $p$ is a positive integer, $a$ is a complex-valued number, and $\Gamma[a]$ is the gamma function. The hypergeometric function ${}_1F_1 (M;M + K; - x \mathbf{I} )$ is 
\begin{equation}
{}_1F_1 \left(M; M + K; - x{\mathbf{I}} \right) = \sum_{k=0}^{\infty} \sum_{\kappa} {\left[ M \right]_{\kappa} \over \left[ M + K \right]_{\kappa}} {C_{\kappa}\left(- x \mathbf{I}\right) \over k!}.
\label{eq:hypergeometric}
\end{equation}
The details of $[M]_{\kappa}$ and $C_{\kappa}(- x \mathbf{I})$ in (\ref{eq:hypergeometric}) can be found in \cite{Ratnarajah_Eig_Wishart}. Based on (\ref{eq:eigen_largest}), the probability of (\ref{eq:convergence_condition3}) can be directly derived. However, (\ref{eq:eigen_largest}) includes  the summation of infinite terms in (\ref{eq:hypergeometric}) which has extreme complexity, so it cannot provide a closed-form convergence condition of (\ref{eq:convergence_condition3}) in terms of $\alpha$. 

Fortunately, based on (\ref{eq:eigenvalues}), the condition of (\ref{eq:convergence_condition3}) changes to 
\begin{equation}
M \left(1 + {1 \over \sqrt{\alpha}} \right)^2 < 2M.
\label{eq:convergence_condition4}
\end{equation} 
Based on (\ref{eq:convergence_condition4}), a high probability convergence condition in terms of $\alpha$ is derived as 
\begin{align}
\alpha > {1 \over \left(\sqrt{2} - 1 \right)^2} \approx 5.83.
\label{eq:convergence_condition5}
\end{align}
With (\ref{eq:convergence_condition5}), the maximum possible number of $K$ can be found for a specific number of $M$ to achieve a very high probability of convergence for (\ref{eq:neumann_condition}). 

\begin{figure}[!t]
\centering \includegraphics[width = 1.0\linewidth]{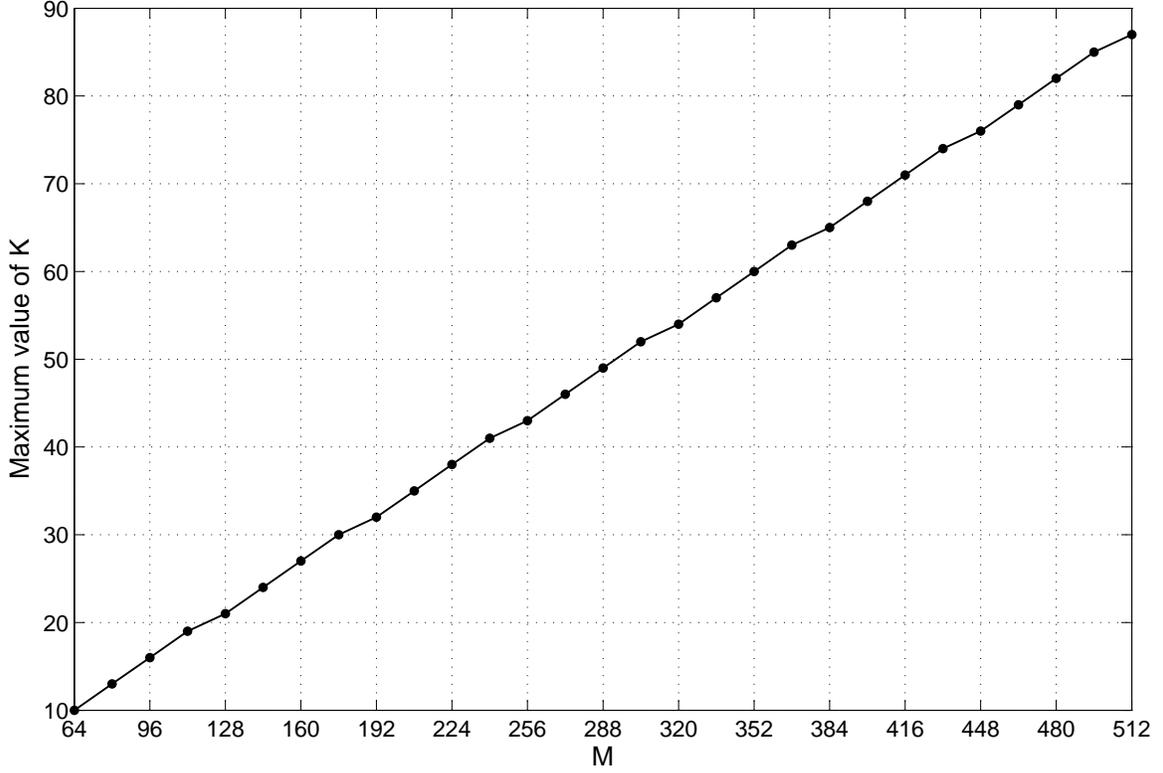}
\caption{The maximum $K$ values for different $M$ values that satisfy (\ref{eq:convergence_condition5})}
\label{fig:k_convergence}
\end{figure}

\begin{figure}[!t]
\centering \includegraphics[width = 1.0\linewidth]{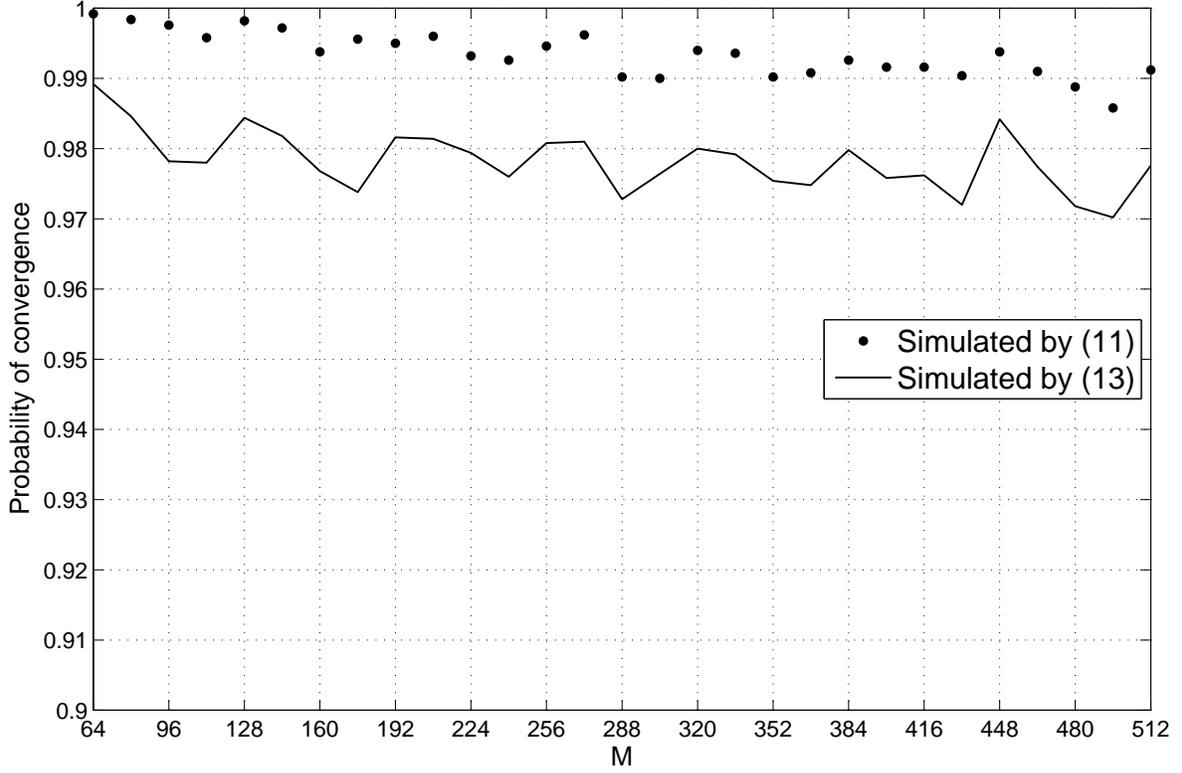}
\caption{Convergence probability values of (\ref{eq:neumann_condition}) for the $K$ values in Fig. \ref{fig:k_convergence}} 
\label{fig:k_convergence_prob}
\end{figure}

Fig. \ref{fig:k_convergence} illustrates the maximum values of $K$ corresponding to $M$ values that vary from $64$ to $512$ based on the convergence condition (\ref{eq:convergence_condition5}). With these $K$ values, the simulated convergence probability values of (\ref{eq:neumann_condition}) based on the accurate condition (\ref{eq:convergence_condition1}) and the approximated condition (\ref{eq:convergence_condition3}) are shown in Fig. \ref{fig:k_convergence_prob}. The results indicate that they provide close probability with (\ref{eq:convergence_condition1}) being slightly better 
in massive MIMO systems with large $M$. The results verify that (\ref{eq:convergence_condition3}) is an acceptable approximation of (\ref{eq:convergence_condition1}). Furthermore, the results show that the condition (\ref{eq:convergence_condition5}) in terms of $\alpha$ can offer high convergence probability for (\ref{eq:neumann_condition}). Table \ref{table:convergence} summarizes the typical $M$ values of massive MIMO systems with their corresponding maximum values of $K$ and the convergence probability values of (\ref{eq:neumann_condition}). Note that (\ref{eq:convergence_condition5}) does not ensure fast convergence of (\ref{eq:neumann_condition}), so a more strict $\alpha$ condition for $\mathbf{G}$ being a DDM is studied in the next section.

\begin{table}[!t]
\normalsize
\renewcommand{\arraystretch}{1.1}
\caption{Typical $M$ values with their associated maximum values of $K$ and convergence probability values}
\centering
\begin{tabular}{|c|c|c|c|c|}
\hline
$M$ & $64$ & $128$ & $256$ & $512$ \\  
\hline
$K$ & $10$ & $21$ & $43$ & $87$ \\ 
\hline
Probability of (\ref{eq:neumann_condition}) & $0.999$ & $0.998$ & $0.995$ & $0.991$ \\ \hline
\end{tabular}
\label{table:convergence}
\end{table}
\section{Diagonally Dominant and $\alpha$} \label{sec:dominant}
Let $\mathbf{h}_k$ denote the $k$th column vector of the $M \times K$ channel matrix $\mathbf{H}$. Then, $\mathbf{h}_k$ represents the $M$-dimensional channel vector for the $k$th user. Hence, the elements of the $K\times K$ matrix $\mathbf{G} = \mathbf{H}^{\mathrm{H}} \mathbf{H}$ is calculated as
\begin{align}
\left\lbrace
\begin{array}{ll}
g_{ii} = \left\|\mathbf{h}_i \right\|^2_2, & i = 1, \ldots, K, \\
g_{ij} = \mathbf{h}_i^{\mathrm{H}} \mathbf{h}_j, & j = 1, \ldots, K, \,\, j \neq i.
\end{array}
\right.
\label{eq:row}
\end{align}
As mentioned in Section \ref{sec:convergence}, the diagonal elements $g_{ii}$ approach to $M$ when the number of $M$ is large. As a result, the requirement (\ref{eq:dominant_condition1}) in Section \ref{sec:basis} for $\mathbf{G}$ being a DDM can be approximated as
\begin{equation}
\Delta_i = \sum_{j\neq i}\left|r_{ij}\right| < 1,\forall i,
\label{eq:dominant_condition2}
\end{equation}
where $r_{ij}$ is the normalized correlation coefficient between $\mathbf{h}_i$ and $\mathbf{h}_j$ defined as
\begin{equation}
r_{ij} = {\mathbf{h}_i^{\mathrm{H}} \mathbf{h}_j \over \left\| \mathbf{h}_i  \right\|_2 \left\| \mathbf{h}_j  \right\|_2} \approx {\mathbf{h}_i^{\mathrm{H}} \mathbf{h}_j \over M}.
\label{eq:normalized_corr}
\end{equation}

Let $x = |r_{ij}|$. Note that the Probability Density Function (PDF) of $x$ was derived in \cite{Zhu_Grassmann_arXiv} as
\begin{align}
f\left( x \right) = 2\left( M -1 \right) x \left(1-x^2\right)^{M-2}, & \,\, 0\leq x \leq 1.
\label{eq:pdf}
\end{align}
Hence, the mean of $x$ is 
\begin{align}
\mathrm{E}\left( x \right) = \int_0^1 x f\left( x \right) \mathrm{d}x 
= \left( M -1 \right) \mathrm{B}\left(1.5, M-1\right),
\label{eq:mean}
\end{align}
where $\mathrm{B}(a,b)$ with $a$ and $b$ being complex-valued numbers is the beta function defined as 
\begin{align}
\mathrm{B}\left(a, b\right) = \int_0^1 t^{a-1}\left(1-t\right)^{b-1} \mathrm{d}t, &\,\, \Re \left\lbrace a \right\rbrace, \Re \left\lbrace b \right\rbrace > 0. 
\label{eq:beta}
\end{align}
Although (\ref{eq:mean}) provides the values of $\mathrm{E}(x)$, since the number of $K$ is not large enough, $\Delta_i$ in (\ref{eq:dominant_condition2}) can be larger than $(K-1)\mathrm{E}(x)$. However, $\Delta_i$ has a high probability being smaller than $(K-1)[\mathrm{E}(x) + \delta(x)]$ where $\delta(x)$ denotes the standard deviation of $x$, which is
\begin{align}
\delta\left(x \right) = \sqrt{\mathrm{E}\left( x^2 \right) - \mathrm{E}\left( x \right)^2}, 
\label{eq:sd}
\end{align}
with
\begin{align}
\mathrm{E}\left( x^2 \right) = \int_0^1 x^2 f\left( x \right) \mathrm{d}x 
= \left( M -1 \right) \mathrm{B}\left(2, M-1\right).
\label{eq:mean_two}
\end{align}
Therefore, the condition (\ref{eq:dominant_condition2}) can be approximated as
\begin{equation}
\left(K-1\right) \left[\mathrm{E}\left(x\right) + \delta\left(x\right)\right] < 1.
\label{eq:dominant_condition3}
\end{equation}
Based on (\ref{eq:dominant_condition3}), a high probability condition for the $\mathbf{G}$ matrix being a DDM in terms of $\alpha$ is derived as
\begin{equation}
\alpha > {M \left[\mathrm{E}\left(x\right) + \delta\left(x\right)\right] \over \mathrm{E}\left(x\right) + \delta\left(x\right) + 1 }.
\label{eq:dominant_condition4}
\end{equation}
With (\ref{eq:dominant_condition4}), the maximum possible number of $K$ can be found for a specific number of $M$ to achieve a very high probability for $\mathbf{G}$ being a DDM. 

\begin{figure}[!t]
\centering \includegraphics[width = 1.0\linewidth]{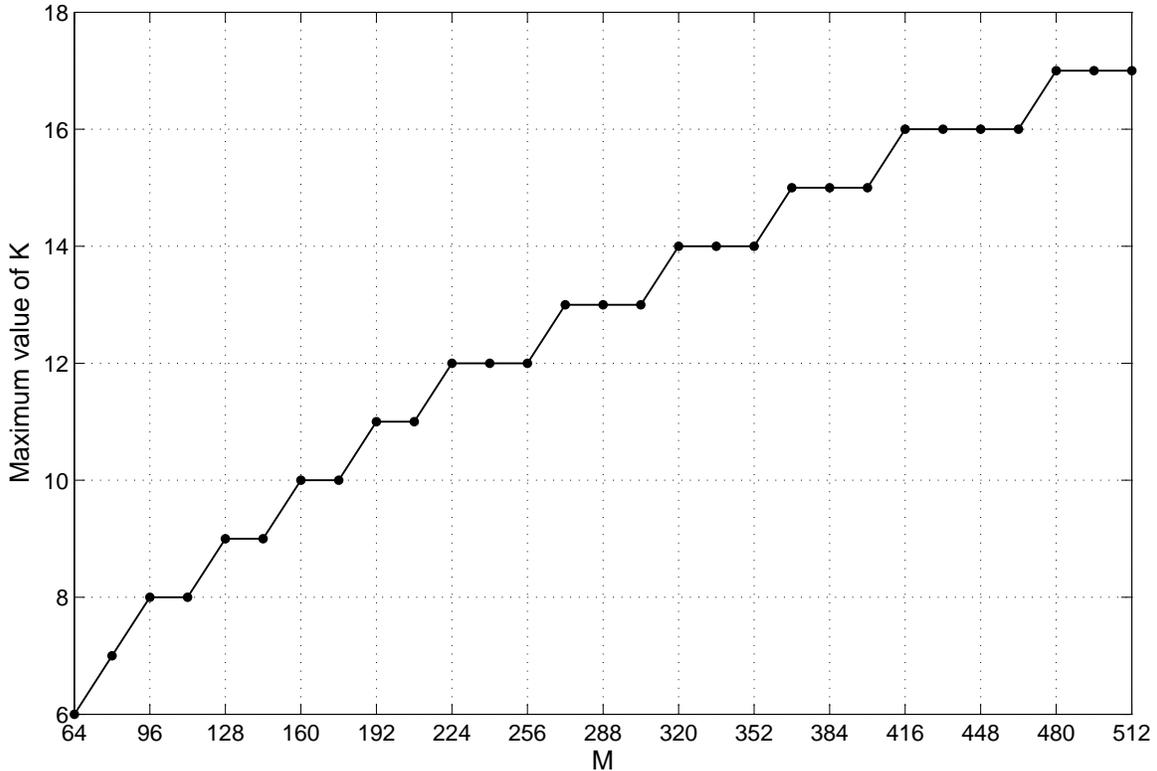}
\caption{The maximum $K$ values for different $M$ values that satisfy (\ref{eq:dominant_condition4})}
\label{fig:k_dominant}
\end{figure}

\begin{figure}[!t]
\centering \includegraphics[width = 1.0\linewidth]{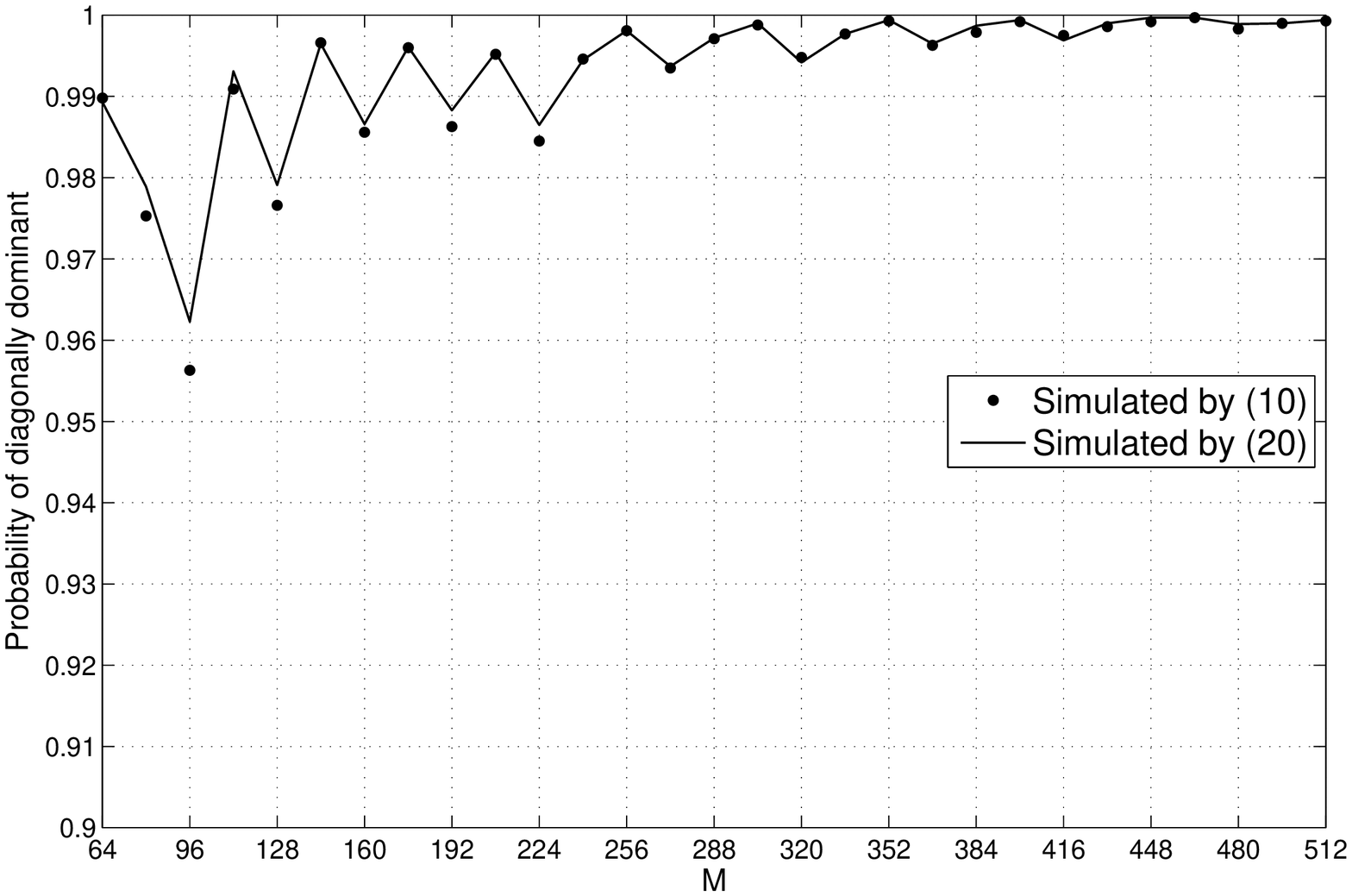}
\caption{Diagonally dominant probability values for the $K$ values in Fig. \ref{fig:k_dominant}} 
\label{fig:k_prob_dominant}
\end{figure}

Fig. \ref{fig:k_dominant} shows the maximum values of $K$ corresponding to $M$ values that vary from $64$ to $512$ based on the diagonally dominant condition (\ref{eq:dominant_condition4}). With these $K$ values, the simulated DDM probability based on the definition (\ref{eq:dominant_condition1}) and the approximated condition (\ref{eq:dominant_condition2}) are illustrated in Fig. \ref{fig:k_prob_dominant}. The results show that they achieve close probability 
in massive MIMO systems with large $M$. The results verify that (\ref{eq:dominant_condition2}) is a good approximation of (\ref{eq:dominant_condition1}), especially when $M$ is very large. Moreover, the results show that the condition (\ref{eq:dominant_condition4}) in terms of $\alpha$ can offer high DDM probability
. Table \ref{table:dominant} summarizes the typical $M$ values of massive MIMO systems with their corresponding maximum values of $K$ and the diagonally dominant probability values of (\ref{eq:dominant_condition1}). Note that the DDM condition (\ref{eq:dominant_condition4}) is sufficient for the convergence condition (\ref{eq:convergence_condition5}) and leads to quicker convergence, so it is more useful in practice. 

\begin{table}[!t]
\normalsize
\renewcommand{\arraystretch}{1.1}
\caption{Typical $M$ values with their associated maximum values of $K$ and diagonally dominant probability values}
\centering
\begin{tabular}{|c|c|c|c|c|}
\hline
$M$ & $64$ & $128$ & $256$ & $512$ \\  
\hline
$K$ & $6$ & $9$ & $12$ & $17$ \\ 
\hline
Probability of (\ref{eq:dominant_condition1}) & $0.990$ & $0.977$ & $0.998$ & $0.999$ \\ \hline
\end{tabular}
\label{table:dominant}
\end{table}

\section{Error Analysis} \label{sec:error}
Based on (\ref{eq:decompose}) and (\ref{eq:diagonal_choice3}), the NS-based MIA of (\ref{eq:neumann_series}) changes to 
\begin{align}
\mathbf{G}^{-1}_N 
= \sum_{n}^{N-1} \left( - \mathbf{D}^{-1}\mathbf{E} \right)^n \mathbf{D}^{-1}.
\label{eq:neumann_decompoase}
\end{align}
Note that if the convergence condition (\ref{eq:neumann_condition}) is satisfied, $\mathbf{G}^{-1}_{\infty}$ is the exact matrix inverse of $\mathbf{G}$. However, in practice, the number of $N$ cannot be very large. Otherwise, it would cause excessive burden for hardware implementation. In this case, residual error resulted from the NS-based MIA $\mathbf{G}^{-1}_N$ exists. Let the $K$-dimensional vector $\mathbf{s}$ denote the transmitted symbols for the uplink or the downlink. Without loss of generality, $\mathrm{E}(|s_k|^2) = 1$ is assumed, with $k=1,\ldots,K$. Let $\mathbf{Z}=\mathbf{D}^{-1}\mathbf{E}$. Then, the Mean Square Error (MSE) of the NS-based MIA $\mathbf{G}^{-1}_N$ for the uplink is derived as
\begin{align}
\epsilon^{\mathrm{ul}}_N & = \mathrm{E} \left\lbrace \left\| \left( \mathbf{G}^{-1}_{\infty} - \mathbf{G}^{-1}_{N} \right) \mathbf{H}^{\mathrm{H}} \mathbf{H} \mathbf{s} \right\|^2_2 \right\rbrace \nonumber \\
& = \mathrm{E} \left\lbrace \left\| \mathbf{Z}^N \sum_{n=0}^{\infty} \left(-\mathbf{Z}\right)^n \mathbf{D}^{-1} \mathbf{H}^{\mathrm{H}} \mathbf{H} \mathbf{s} \right\|^2_2 \right\rbrace \nonumber \\
& = \mathrm{E} \left\lbrace \left\| \mathbf{Z}^N \mathbf{G}^{-1}_{\infty} \mathbf{G} \mathbf{s} \right\|^2_2 \right\rbrace \nonumber \\
& = \mathrm{E} \left\lbrace \left\| \mathbf{Z}^N \mathbf{s} \right\|^2_2 \right\rbrace \nonumber \\ 
& = \mathrm{E} \left\lbrace \mathrm{Tr} \left[ \mathbf{Z}^N \mathbf{s} \mathbf{s}^{\mathrm{H}} \left(\mathbf{Z}^N \right)^{\mathrm{H}} \right] \right\rbrace \nonumber \\  
& = \mathrm{E} \left\lbrace \mathrm{Tr} \left[ \mathbf{s} \mathbf{s}^{\mathrm{H}}  \left(\mathbf{Z}^N \right)^{\mathrm{H}} \mathbf{Z}^N \right] \right\rbrace \nonumber \\  
& = \mathrm{Tr} \left\lbrace \mathrm{E} \left[ \mathbf{s} \mathbf{s}^{\mathrm{H}} \right] \mathrm{E} \left[\left(\mathbf{Z}^N \right)^{\mathrm{H}} \mathbf{Z}^N \right] \right\rbrace \nonumber \\  
& = \mathrm{Tr} \left\lbrace \mathbf{I}_K \mathrm{E} \left[\left(\mathbf{Z}^N \right)^{\mathrm{H}} \mathbf{Z}^N \right] \right\rbrace \nonumber \\  
& = \mathrm{E} \left\lbrace \mathrm{Tr} \left[ \left(\mathbf{Z}^N \right)^{\mathrm{H}} \mathbf{Z}^N \right] \right\rbrace \nonumber \\ 
& = \mathrm{E} \left\lbrace \left\| \mathbf{Z}^N \right\|_{\mathrm{F}}^2 \right\rbrace.
\label{eq:mmse_ul}  
\end{align}
Note that for the downlink case, the MSE result is 
\begin{align}
\epsilon^{\mathrm{dl}}_N & = \mathrm{E} \left\lbrace \left\| \mathbf{s}^{\mathrm{T}} \left( \mathbf{G}^{-1}_{\infty} - \mathbf{G}^{-1}_{N} \right) \mathbf{H}^{\mathrm{H}} \mathbf{H}  \right\|^2_2 \right\rbrace \nonumber \\
& = \mathrm{E} \left\lbrace \left\| \mathbf{s}^{\mathrm{T}} \mathbf{Z}^N  \right\|^2_2 \right\rbrace \nonumber \\ 
& = \mathrm{E} \left\lbrace \left\| \mathbf{Z}^N \right\|_{\mathrm{F}}^2 \right\rbrace,
\label{eq:mmse_dl}  
\end{align}
which is the same as (\ref{eq:mmse_ul}). Hence, $\epsilon_N$ is used instead of $\epsilon^{\mathrm{ul}}_N$ and $\epsilon^{\mathrm{dl}}_N$ 
in 
this section below. Since $\epsilon_N$ can be interpreted as the power of the residual interference of ZF precoding or detection, the average Signal-to-Interference Ratio (SIR) for each user 
is calculated as
\begin{align}
\gamma_N = {{\left\|\mathbf{s}\right\|^2 \over K} \over {\epsilon_N \over K}} 
= {K \over \epsilon_N}.
\label{eq:gamma}
\end{align}

\begin{figure}[!t]
\centering \includegraphics[width = 1.0\linewidth]{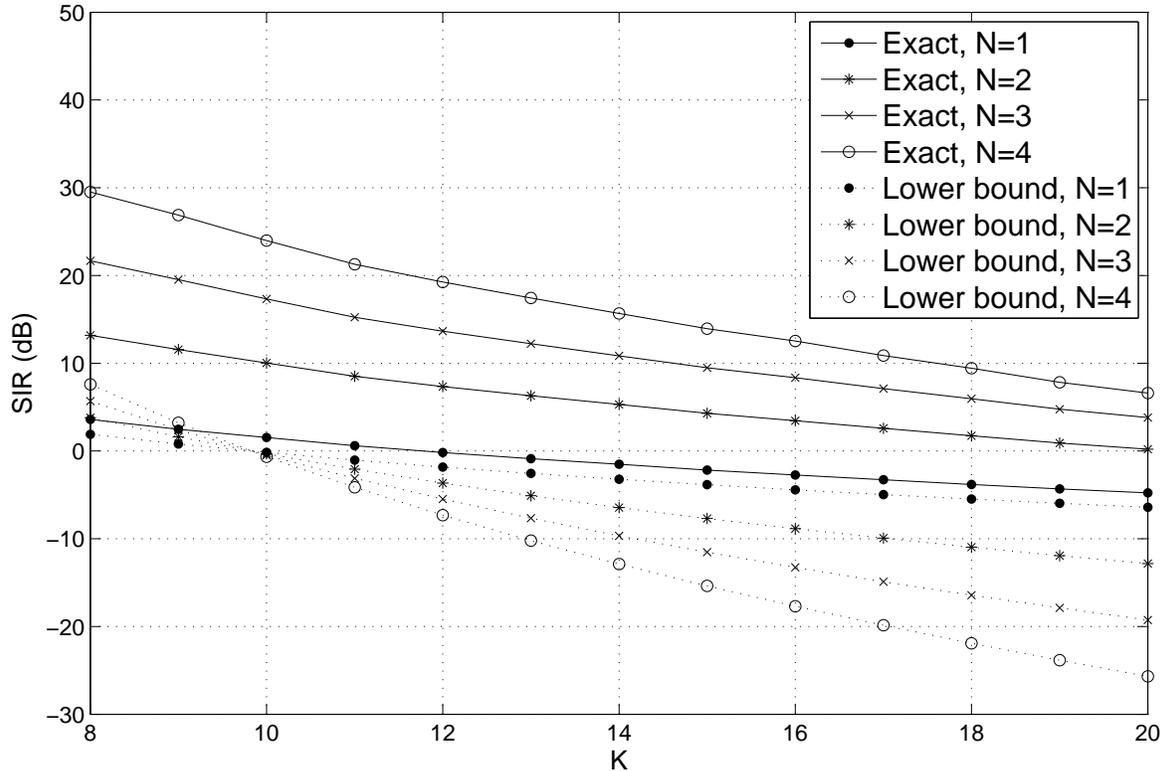}
\caption{Comparison between the exact SIR values and the lower bound values resulting from (\ref{eq:error_ub}) with $M=128$ for different $K$ and $N$ values.}
\label{fig:gamma_lb}
\end{figure}

In \cite{Wu_Neumann}, the MSE $\epsilon_N$ in (\ref{eq:mmse_ul}) and (\ref{eq:mmse_dl}) is upper bounded as
\begin{align}
\epsilon_N 
\leq \left[ (K^2-K)\sqrt{2M(M+1) \over \left(M-1\right)\left(M-2\right)\left(M-3\right)\left(M-4\right)} \right]^N,
\label{eq:error_ub}
\end{align}
with $M>4$. Unfortunately, (\ref{eq:error_ub}) is a very loose upper bound, resulting in a very loose lower bound of $\gamma_N$ in (\ref{eq:gamma}). In Fig. {\ref{fig:gamma_lb}}, the exact SIR values and the lower bound values are compared with $M=128$ for different $K$ and $M$ values. The results show substantial differences when $N>1$, 
which cannot provide sufficient insight for the residual error of the NS-based MIA for $N>1$. Due to this reason, we seek to derive a more accurate approximation of $\epsilon_N$ in this section below.

When $M$ is large, because $\mathbf{D}$ can be approximated as $M\mathbf{I}_K$ as mentioned in Section \ref{sec:convergence}, 
according to (\ref{eq:row}) and (\ref{eq:normalized_corr}), the elements of $\mathbf{Z}$ is approximated as 
\begin{align}
\left\lbrace
\begin{array}{ll}
z_{ii} = 0 & i = 1, \ldots, K, \\
z_{ij} = z_{ji}^* \approx { \mathbf{h}_i^{\mathrm{H}} \mathbf{h}_j \over M}\approx r_{ij}, & j = 1, \ldots,K, \,\, j \neq i.
\end{array}
\right.
\label{eq:elements_zmatrix}
\end{align}
As a result, the PDF of $x=|z_{ij}|$ can be approximated as (\ref{eq:pdf}). Then, a more accurate approximation of $\gamma_N$ can be derived based on (\ref{eq:pdf}).

When $N=1$, the MSE $\epsilon_N$ in (\ref{eq:mmse_ul}) and (\ref{eq:mmse_dl}) changes to
\begin{equation}
\epsilon_1 = \left\| \mathbf{Z} \right\|_F^2 = \sum_{i=1}^K \sum_{j=1, j\neq i} ^K \left| z_{ij} \right|^2    \approx K\left(K - 1 \right)\mathrm{E}\left( x^2 \right).
\label{eq:error1}
\end{equation}
Since $\mathrm{E}(x^2)$ has been derived as (\ref{eq:mean_two}), the term  $\epsilon_1$ in (\ref{eq:error1}) is rewritten as  
\begin{equation}
\epsilon_1 \approx K\left(K-1\right) \mathrm{B}_{2,M},
\label{eq:error1_result}
\end{equation}
where $\mathrm{B}_{a,M}$ is defined as 
\begin{equation}
\mathrm{B}_{a,M} = \left( M -1 \right) \mathrm{B}\left(a, M-1\right).
\label{eq:beta_simple}
\end{equation}

When $N=2$, the MSE $\epsilon_N$ in (\ref{eq:mmse_ul}) and (\ref{eq:mmse_dl}) changes to
\begin{equation}
\epsilon_2 = \left\| \mathbf{Z}^2 \right\|_{\mathrm{F}}^2,
\label{eq:error2}
\end{equation}
where the elements in $\mathbf{Y}=\mathbf{Z}^2$ is
\begin{align}
\left\lbrace
\begin{array}{ll}
y_{ii} = \sum\limits_{k = 1,k \neq i}^K \left| z_{ik} \right|^2, & i = 1, \ldots, K, \\
y_{ij} = \sum\limits_{k = 1,k \neq i,j}^K z_{ik} z_{jk}^*, & j = 1, \ldots, K, \,\, j \neq i.
\end{array}
\right.
\label{eq:ymatrix2}
\end{align}
Note that $\|\mathbf{Z}^2\|^2_{\mathrm{F}}$ can be written as a summation of polynomial terms, which can be classified into three categories. The first category includes $(K-1)K$ terms of $| z_{ik} |^4 $ with $i \neq k$. The second category includes $(K-2)(K-1)K$ terms of $| z_{ik}|^2 | z_{il} |^2 $ with $i \neq k \neq l$, as well as $(K-2)(K-1)K$ terms of $| z_{ik}|^2 | z_{jk} |^2 $ with $i \neq j \neq k$. Hence, the total number of terms for the second category is $2(K-2)(K-1)K$. Finally, the third category includes $(K-3)(K-2)(K-1)K$ terms of $z_{ik} z_{jk}^* z_{il}^* z_{jl}$ with $i \neq j \neq k \neq l$.
Because the elements of $\mathbf{H}$ are i.i.d zero-mean unit-variance complex Gaussian random variables, based on (\ref{eq:elements_zmatrix}), the elements of $z_{ij}$ are i.i.d. zero-mean random variables. As a result, the terms of the third category are also i.i.d. zero-mean random variable. Therefore, the sum of the terms of the third category can be approximated as zero. For the terms of the first category, the mean can be calculated based on (\ref{eq:pdf}) as
\begin{equation}
\mathrm{E} \left( x^4 \right) = \int_0^1 x^4 f\left(x \right)\mathrm{d} x = B_{3,M}. 
\label{eq:mean_four}
\end{equation} 
Similarly, the mean of the terms of the second category is can be approximated as 
\begin{equation}
\mathrm{E}\left( x_1^2  x_2^2 \right) = \mathrm{E}\left( x^2 \right)^2 = B_{2,M}^2. 
\label{eq:mean_four_cross}
\end{equation}
Due to (\ref{eq:mean_four}) and (\ref{eq:mean_four_cross}), the term $\epsilon_2$ in (\ref{eq:error2}) is approximated as 
\begin{equation}
\epsilon_2 \approx K\left( K - 1 \right)B_{3,M} + 2\left( K - 2 \right)\left( K - 1 \right)KB_{2,M}^2.
\label{eq:error2_result}
\end{equation}

When $N>2$, the MSE $\epsilon_N$ in (\ref{eq:mmse_ul}) and (\ref{eq:mmse_dl}) can be derived with the similar method applied by $N=2$. The results of $N=3$ and $N=4$ are directly provided below as
\begin{align}
\epsilon_3 & \approx \left( K - 2\right) \left( K - 1 \right) K\left( 5K - 8 \right) B_{2,M}^3  \nonumber \\ & \,\,\,\,\,\, + \left( 2K - 3 \right)\left( K - 1 \right)K B_{3,M} B_{2,M}, 
\label{eq:error3_result}
\end{align}
and
\begin{align}
\epsilon_4 & \approx \left( 2K - 3 \right)\left( K - 1 \right)KB_{3,M}^2 \nonumber \\ 
& \,\,\,\,\,\, + \left( 2K - 3 \right)^2 \left( K - 1 \right)^2 KB_{4,M} B_{2,M} \nonumber \\ 
& \,\,\,\,\,\, + \left( K - 2 \right)\left( K - 1 \right)^2 K^2 B_{2,M}^4. 
\label{eq:error4_result}
\end{align}

\begin{figure}[!t]
\centering \includegraphics[width = 1.0\linewidth]{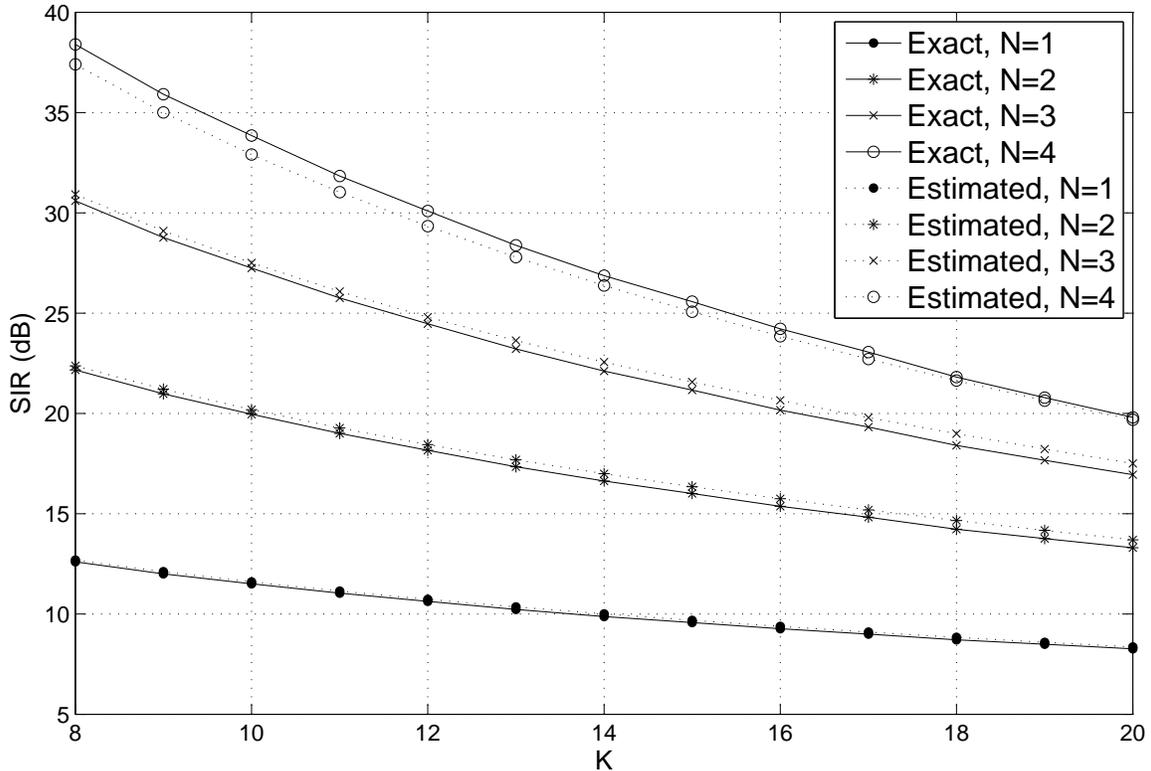}
\caption{Comparison between the exact and estimated SIR values with $M=128$ for different $K$ and $N$ values.}
\label{fig:gamma_est_128}
\end{figure}

\begin{figure}[!t]
\centering \includegraphics[width = 1.0\linewidth]{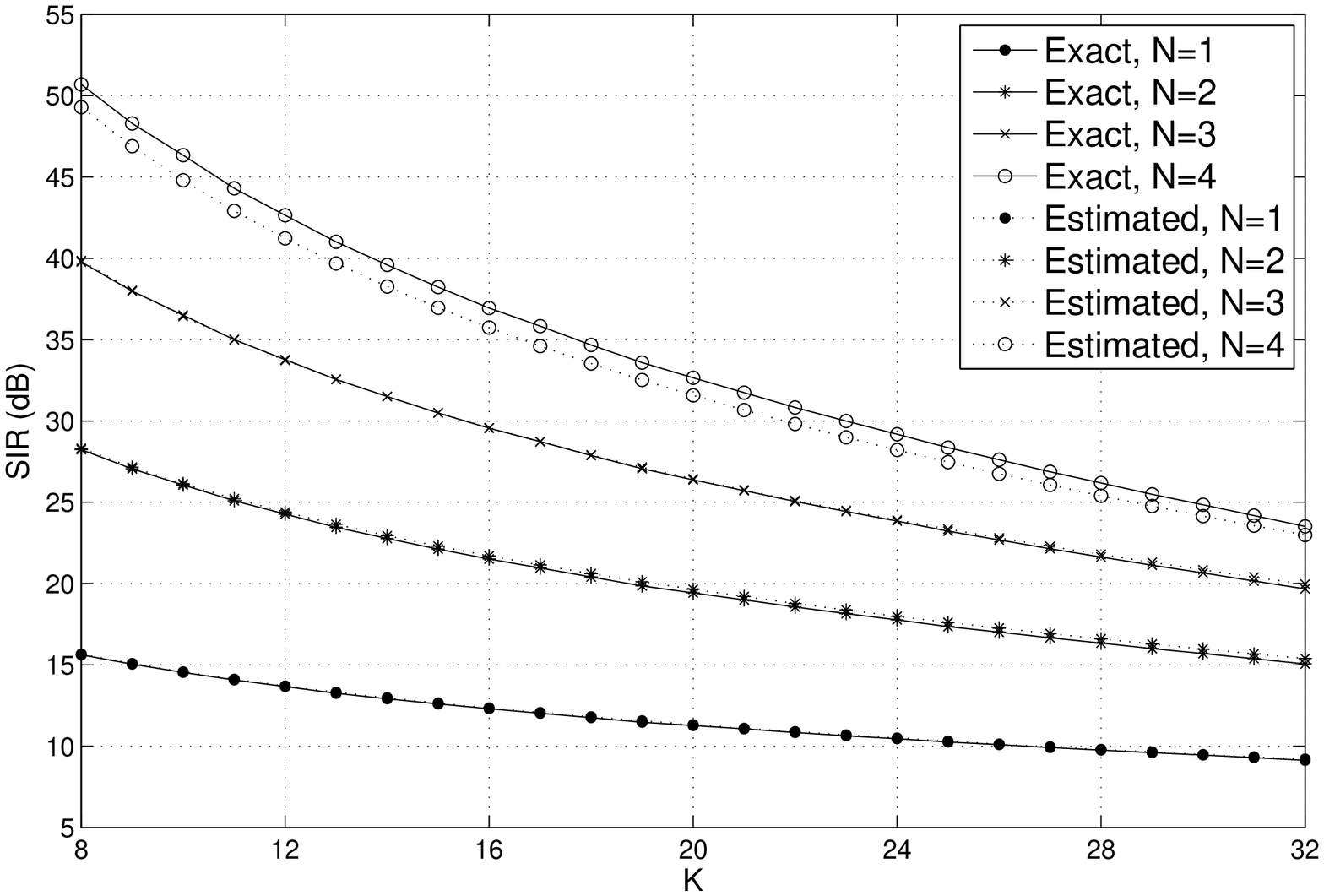}
\caption{Comparison between the exact and estimated SIR values with $M=256$ for different $K$ and $N$ values.}
\label{fig:gamma_est_256}
\end{figure}

\begin{figure}[!t]
\centering \includegraphics[width = 1.0\linewidth]{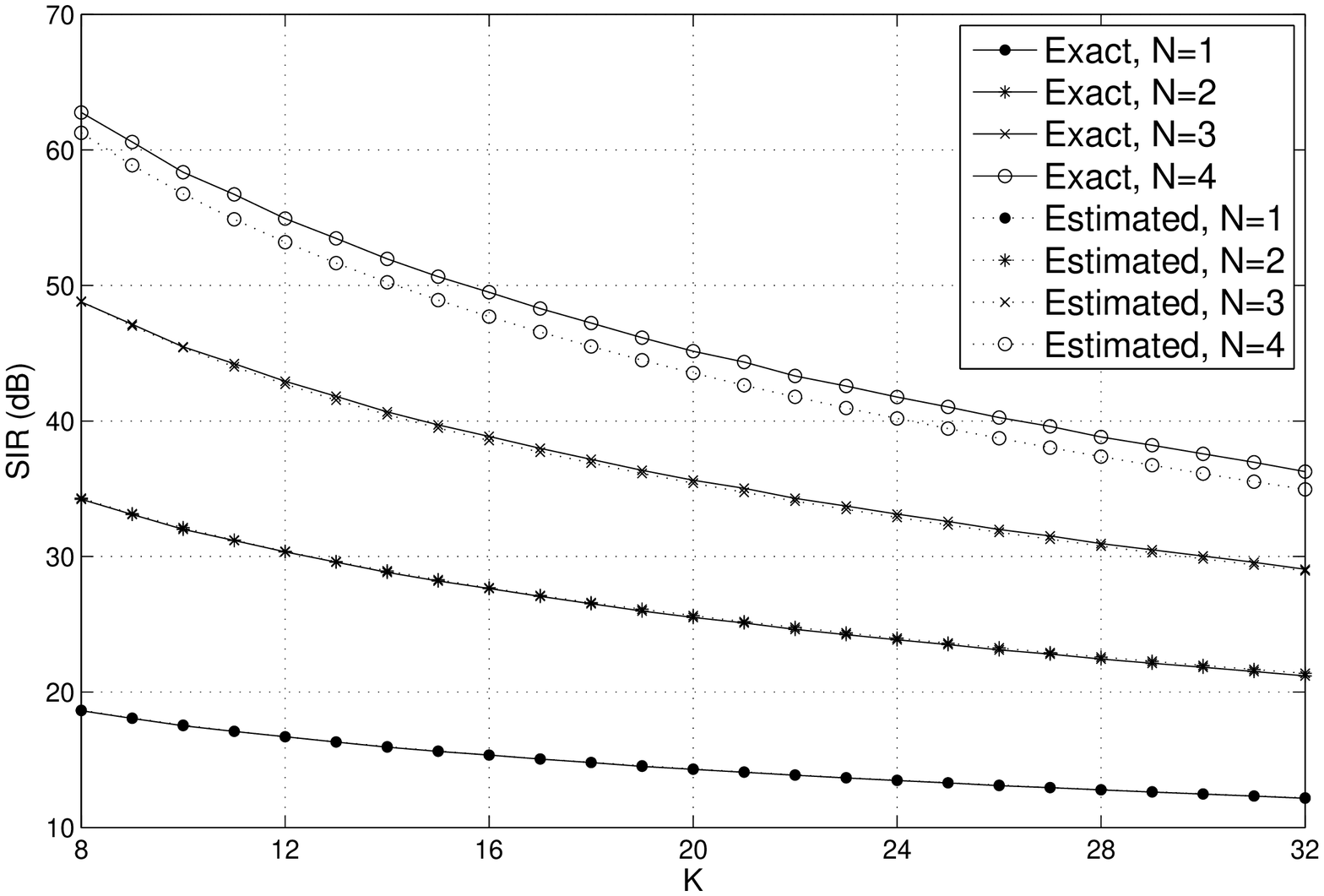}
\caption{Comparison between the exact and estimated SIR values with $M=512$ for different $K$ and $N$ values.}
\label{fig:gamma_est_512}
\end{figure}

With the estimated residual error formulas (\ref{eq:error1_result}), (\ref{eq:error2_result})-(\ref{eq:error4_result}), the estimated SIR formulas can be easily derived according to (\ref{eq:gamma}). Fig. \ref{fig:gamma_est_128}-\ref{fig:gamma_est_512} compare the exact and estimated SIR values for different $K$ and $N$ values, with $M=128$, $M=256$, and $M=512$ respectively. The results show that the estimated SIR values are very close to the exact SIR values, which verifies the high accuracy of SIR estimation formulas based on (\ref{eq:error1_result}), (\ref{eq:error2_result})-(\ref{eq:error4_result}). 
\section{Discussions} \label{sec:discussion}
In massive MIMO systems, $\alpha$ is commonly considered to be very large to offer good performance \cite{Hoydis_massive_MIMO,Rusek_massive_MIMO_overview}, e.g., $\alpha > 10$. Hence, the convergence condition (\ref{eq:convergence_condition5}), i.e., $\alpha >5.83$, derived in Section \ref{sec:convergence} is generally satisfied for massive MIMO systems. Note that the convergence probability values provided in Fig. \ref{fig:k_convergence} and Table \ref{table:convergence}, which are already close to $1$, correspond to the smallest $\alpha$ values that satisfy (\ref{eq:convergence_condition5}). Hence, the convergence probability values for massive MIMO systems are not lower than the values provided in Fig. \ref{fig:k_convergence} and Table \ref{table:convergence}. Therefore, the convergence of NS-based MIA is guaranteed 
so that it 
is a valid method for massive MIMO systems, and its accuracy can be improved by increasing $N$.

As mentioned at the end of Section \ref{sec:convergence}, the convergence condition (\ref{eq:convergence_condition5}) does not guarantee quick convergence of (\ref{eq:neumann_condition}). With the diagonally dominant condition (\ref{eq:dominant_condition4}) derived in Section \ref{sec:dominant}, however, the NS-based MIA can achieve good accuracy with quick convergence, i.e., a small $N$ can offer a sufficiently good MIA. Otherwise, with the same $N$ value, violating (\ref{eq:dominant_condition4}) results in performance loss for the ZF decoding or detection employing the NS-based MIA. Take the simulation results provided in \cite{Wu_Neumann} as examples, with $M=128$ and $N=3$, the choice of $K=4$ satisfying (\ref{eq:dominant_condition4}) achieves close performance to the exact inverse, while the choice of $K=12$ violating (\ref{eq:dominant_condition4}) suffers huge performance loss. However, (\ref{eq:dominant_condition4}) requires very small $\alpha$ values, and $\alpha$ becomes smaller as $M$ increases, which can be seen from Table \ref{sec:dominant}. The strict requirement of $\alpha$ may reduce the spatial multiplexing advantage of massive MIMO systems, i.e., at most $K=17$ users can be served by $M=512$ antennas. To relieve this issue, one comprised choice is to apply an $\alpha$ slightly higher than (\ref{eq:dominant_condition4}) with slightly larger $N$ of the NS-based MIA, depending on the hardware capability.

The SIR discussed in Section \ref{sec:error} reflects the performance error floor for ZF precoding or detection employing practical NS-based MIA in massive MIMO systems. The performance error floor decides the best performance that the ZF precoding or detection employing the NS-based MIA can achieves
. As a result, with $M$, $K$, and $N$, the best achievable performance can be easily estimated based on (\ref{eq:error1_result}), (\ref{eq:error2_result})-(\ref{eq:error4_result}). In addition, since larger $N$ causes higher hardware implementation complexity, with $M$, $K$, and the target performance, the smallest choice of $N$ that can offer sufficiently good performance can be determined to relieve the complexity. Note that a revised form of (\ref{eq:neumann_series}) was provided in \cite{Prabhu_Neumann} as
\begin{align}
\mathbf{G}^{-1}_{N} & \approx \sum_{n=0}^{N-1} \left( \mathbf{I}_K - \mathbf{\Theta} \mathbf{G} \right)^n \mathbf{\Theta} \nonumber \\
& = \prod_{l=0}^{L-1} \left[ \mathbf{I}_K + \left( \mathbf{I}_K - \mathbf{\Theta} \mathbf{G} \right)^{2^l} \right] \mathbf{\Theta},
\label{eq:neumann_series_alternative}
\end{align}
where $L$ is a positive integer with $N=2^L$. Hence, $L=1$, $L=2$, and $L=3$ of the alternative expression (\ref{eq:neumann_series_alternative}) correspond to $N=2$, $N=4$, and $N=8$ of the regular expression (\ref{eq:neumann_series}) respectively. As a result, with the alternative expression (\ref{eq:neumann_series_alternative}), after the choice of $N=4$, the NS-based MIA with the choice of $N=8$ can be quickly calculated. Therefore, if the choice of $N=4$ is not good enough based on the estimation formula (\ref{eq:error4_result}), the choice of $N=8$ can be directly selected based on (\ref{eq:neumann_series_alternative}). Furthermore, note that the complexity of the NS-based MIA with the choice of $N>3$ is considered to be $\mathrm{O}(K^3)$ in \cite{Wu_Neumann}, which loses the complexity advantage over the exact matrix inverse of $\mathrm{O}(K^3)$. In fact, however, the NS-based MIA can be implemented as a series of cascaded matched filter so that the complexity can be reduced to $\mathrm{O}(K^2)$, as discussed in \cite{Rusek_massive_MIMO_overview}. In this way, the NS-based MIA still has the complexity advantage over the exact inverse even with the choice of $N=8$. 
\section{Conclusions} \label{sec:conclusions}
In this paper, three issues related to the practical application of the NS-based MIA in massive MIMO systems are addressed. Firstly, $\alpha > 5.83$ as in (\ref{eq:convergence_condition5}) is offered for the NS-based MIA to achieve very high convergence probability. In other words, with the number of BS antennas $M$, the maximum number of served users $K$ for the NS-based MIA to be a valid method in massive MIMO systems can be determined. Then, a tighter condition (\ref{eq:dominant_condition4}) is provided for $\mathbf{G}$ to be a DDM in very high probability, resulting in a good NS-based MIA with a small number of $N$. This means that given the number of BS antennas $M$, the maximum number of served users $K$ for the NS-based MIA to achieve good performance and quick convergence for ZF decoding or detection cab be determined. Finally, by approximation error analysis, residual error estimation formulas (\ref{eq:error1_result}), (\ref{eq:error2_result})-(\ref{eq:error4_result}) with very high accuracy are derived for practical $N$ values, which can be applied to estimate the error floor caused by the NS-based MIA. Thus, given the number of BS antennas $M$, the number of served users $K$, and the number of terms employed by the NS-based MIA $N$, highly accurate estimation of the SIR caused by the NS-based MIA can be obtained. These results offer useful guidelines for practical application of the NS-based MIA in massive MIMO systems. 

\bibliographystyle{IEEEtran}
\bibliography{IEEEabrv,Mybib}
\end{document}